\documentclass[epj]{svjour}

\usepackage{graphicx}
\usepackage{fancyhdr}
\def\makeheadbox{{
 }}
\def\mytitle{My title} 
\def\myauthors{My name}  
\def\mytype{My type of session}
\def\mysession{My session}

\def\mytitle{Search for single top production at CDF} 
\def\myauthors{Dominic Hirschb\"uhl (for the CDF Collaboration}    
\def\mytype{Contributed Talk}    
\def\mysession{Colliders - Higgs Phenomenology}

\def\EtMiss  {E_\mathrm{T}\!\!\!\!\!\!\!/ \ \; }
\def\EtMissVec {{E}_\mathrm{T}\!\!\!\!\!\!\!/ \;\;}

\begin{document}
\title{Search for single top production using multivariate analyses at CDF}
\author{Dominic Hirschb\"uhl (for the CDF Collaboration) 
}                     
%
%
\institute{Institut f\"ur Experimentelle Kernphysik \\
 Universit\"at Karlsruhe, Wolfgang-Gaede-Stra{\ss}e 1, 76131 Karlsruhe}
%
\date{}
\abstract{This article reports on recent searches for single-top-quark 
production by the CDF collaboration at the Tevatron using a data set 
that corresponds to an integrated luminosity of $955\,\mathrm{pb^{-1}}$.
Three different analyses techniques are employed, one using 
likelihood discriminants, one neural networks and one matrix elements.
The sensitivity to single-top production at the rate predicted by
the standard model ranges from 2.1 to $2.6\,\sigma$.
While the first two analyses observe a deficit of single-top like
events compared to the expectation, the matrix element method observes
an excess corresponding to a background fluctuation of $2.3\,\sigma$.  
The null results of the likelihood and neural network analyses 
translate in upper limits on the cross section of $2.6\,\mathrm{pb}$ 
for the $t$-channel production
mode and $3.7\,\mathrm{pb}$ for the $s$-channel mode
at the 95\% C.L. The matrix element result corresponds to a measurement
of $2.7^{+1.5}_{-1.3}\;\mathrm{pb}$ for the combined $t$- and
$s$-channel single-top cross section. \\ 
In addition, CDF has searched for non-standard model
production of single-top-quarks via the $s$-channel exchange of
a heavy $W^\prime$ boson. 
No signal of this process is found resulting in lower mass limits
of $760\,\mathrm{GeV}/c^2$ in case the mass of the right-handed 
neutrino is smaller than the mass of the right-handed $W^\prime$
or $790\,\mathrm{GeV}/c^2$ in the opposite case.
} 
\maketitle
\section{Introduction}
In $p\bar{p}$ collisions at the Tevatron top quarks are mainly produced in 
pairs via the strong force. However, the standard model also predicts the 
production of single top-quarks by the weak interaction via the $s$- or 
$t$-channel exchange of an off-shell $W$ boson. 
While early Run II searches by the CDF and D\O~collaborations, based on 
data sets corresponding to $162\,\mathrm{pb^{-1}}$, 
$230\,\mathrm{pb^{-1}}$ or $695\,\mathrm{pb^{-1}}$ of integrated luminosity, 
did not find evidence for single-top 
production~\cite{cdfRunII,dzeroRunIIearly,cdfWinter06}, one of the latest 
D\O~analyses~\cite{dzeroEvidence} using data with 
$L_\mathrm{int}=1\,\mathrm{fb^{-1}}$ observes an excess of single-top-like
events of $3.4\,\sigma$.   
 
The single-top production cross section is predicted to 
$\sigma_\mathrm{s+t}=2.9\pm 0.4\,\mathrm{pb}$ for a top quark mass of 
175 GeV/$c^{2}$~\cite{theo} which is about 40\% of the top-antitop 
pair production cross section. The precise measurement of the production cross section allows the direct extraction of the Cabbibo-Kobayashi-Maskawa matrix element $|V_{tb}|$. Moreover, the search for single top also probes exotic models beyond the Standard Model. New physics, like flavor-changing neutral currents or heavy $W'$ bosons, could alter the observed production rate~\cite{exotic}.

The main obstacle in finding
single-top is however not the production rate of the signal but the large 
background rate. 
After all selection requirements are imposed, the signal to
background ratio is approximately 1/16. This challenging,
background-dominated dataset is the main motivation for using
multivariate techniques.

Furthermore, single-top analyses are an important stepping stone for the Higgs 
boson search in the $WH$ channel, since the two processes have the
same experimental signature. Understanding the background rates and 
background composition in the W+jets sample using single-top techniques
will be a prerequisite  for a successful WH analysis.

\section{Standard Model Searches}
In this article we present three new CDF searches for standard model 
single-top production and one new search for a $W^\prime$ boson decaying
into $t\bar{b}$. All four analyses are based on the same event selection
and use the same Run II data set corresponding to an integrated luminosity of
$955\,\mathrm{pb^{-1}}$.
The event selection exploits the kinematic features of the signal
final state, which contains a top quark, a bottom quark, 
and possibly additional light quark jets. 
To reduce multi-jet backgrounds, the $W$ originating from the top
quark is required to have decayed leptonically. 
One therefore demands a single, isolated high-energy
electron or muon ($E_{T}(e)>20$ GeV, or $P_{T}(\mu)>20$ GeV/$c$) 
and large missing transverse energy from the undetected 
neutrino $\EtMiss$\/$>$25 GeV. 
Electrons are measured in the central and in the forward 
calorimeter, $|\eta|<2.0$.
To further suppress events in which no real $W$ boson is produced, called 
non-$W$ background, additional 
cuts are applied. The cuts are based on the assumption that these events do not 
produce $\EtMiss$ by nature but due to lost or mismeasured jets. Therefore, one 
would expect small $\EtMiss$ and small values of the angle $\Delta \phi$ between 
$\EtMissVec$ and a jet.
We further reject dilepton events 
from $Z$ decays by requiring the dilepton mass to
be outside the range: 76 GeV/$c^{2}<M_{\ell \ell}<106$ GeV/$c^{2}$. 

The remaining backgrounds belong to the following categories: 
$Wb\bar{b}$, $Wc\bar{c}$, $Wc$, mistags (light quarks 
misidentified as heavy flavor jets), non-$W$ and diboson $WW$, $WZ$, and $ZZ$.
We remove a large fraction of the backgrounds by demanding exactly 
two jets with $E_{T}>15$ GeV and $|\eta|<2.8$ be present in the event. 
At least one of these two jets should be tagged as a $b$ quark jet 
by using displaced vertex information from the silicon vertex 
detector (SVX).
The numbers of expected and observed events are listed 
in table~\ref{tab:nnexpect}. 

\begin{table}
\caption{Expected number of signal and background events 
and total number of events observed in 955 pb$^{-1}$ in the
CDF single-top dataset.}
\label{tab:nnexpect}
\begin{tabular}{lc|lc}
\hline\noalign{\smallskip}
Process           & $N$ events  &  Process  & $N$ events \\
\noalign{\smallskip}\hline\noalign{\smallskip}
$W+b\bar{b}$      & $170.9\pm 50.7$ & non-$W$    & $26.2\pm 15.9$ \\
$W+c\bar{c}$      & $63.5\pm 19.9$  & $t\bar{t}$ & $58.4\pm13.5$  \\
$Wc$              & $68.6\pm 19.0$  & Diboson    & $13.7\pm1.9$   \\
Mistags           & $136.1\pm 19.7$ & $Z$ + jets & $11.9\pm4.4$   \\
\noalign{\smallskip}\hline\noalign{\smallskip}
\multicolumn{2}{l}{Total background}   & \multicolumn{2}{c}{$549.3\pm 95.2$} \\
\noalign{\smallskip}\hline\noalign{\smallskip}
$t$-channel              & $22.4\pm3.6$ & $s$-channel & $15.4\pm2.2$ \\
\noalign{\smallskip}\hline\noalign{\smallskip}
\multicolumn{2}{l}{Total prediction} & \multicolumn{2}{c}{$587.1\pm 96.6$} \\
\noalign{\smallskip}\hline\noalign{\smallskip}
\multicolumn{2}{l}{Observed}  & \multicolumn{2}{c}{644} \\
\noalign{\smallskip}\hline
\end{tabular}
\end{table}

\subsection{Likelihood Discriminant Analysis}
\label{sec:cdfll}
One multi-variate analysis uses likelihood discriminants to combine several
variables to a discriminant to separate single-top events from background events.
One likelihood discriminant is defined for the $t$-channel, one for the 
$s$-channel search. Seven or six variables are used, respectively.
The likelihood functions are constructed by first forming histograms of each 
variable. The histograms are produced separately for signal and several background
processes. The histograms are normalized such that the sum of their bin contents
equals 1. For one variable the different processes are combined by computing the
ratio of signal and the sum of the background histograms. These ratios are 
multiplicatively combined to form the likelihood discriminant. 

\begin{figure}[!th]  
\hspace*{0.03\textwidth} {\small (a)}\\
\begin{center}
\includegraphics[width=0.40\textwidth]{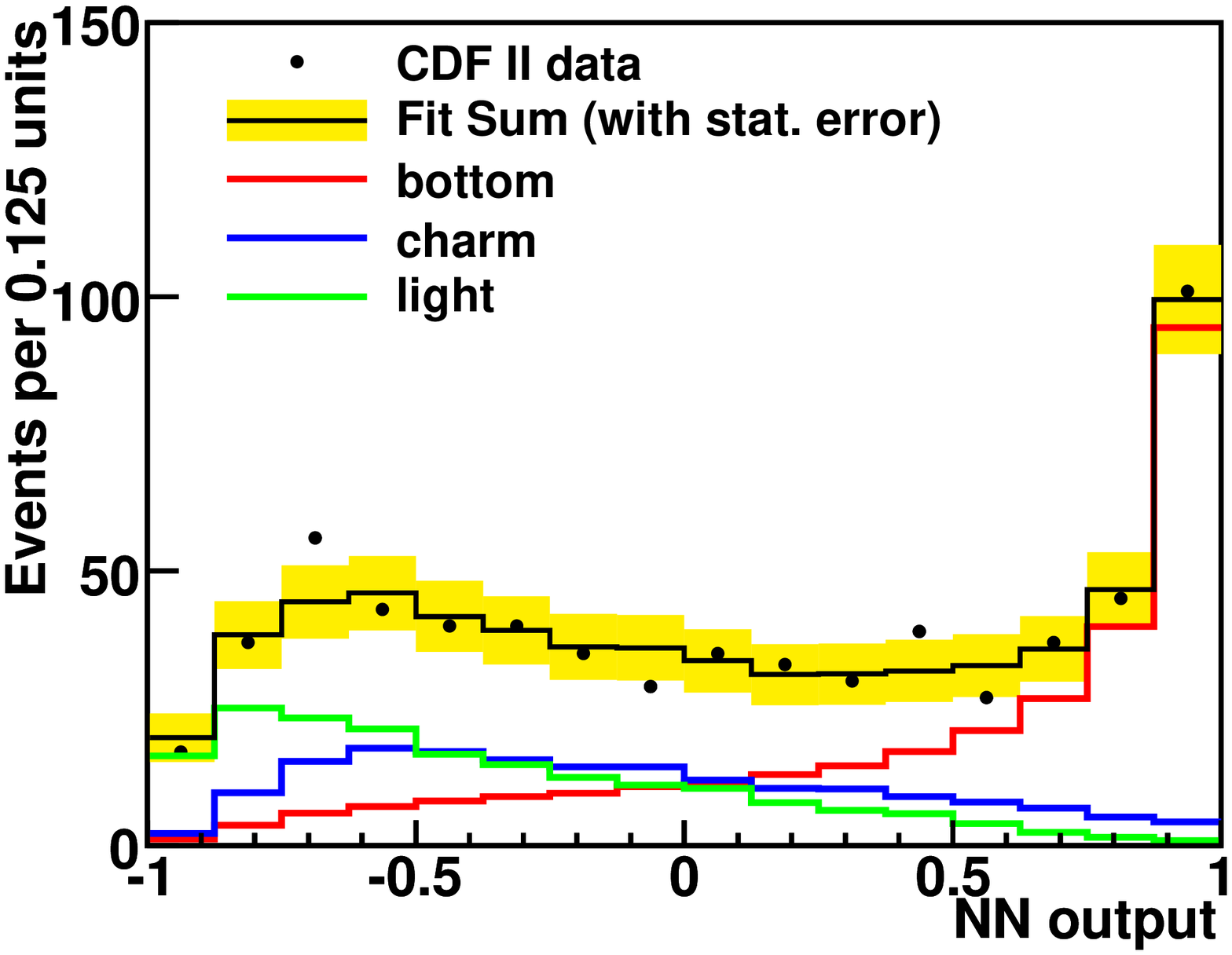}\\
\end{center}
\hspace*{0.03\textwidth} {\small (b)}\\
\begin{center}
\includegraphics[width=0.40\textwidth]{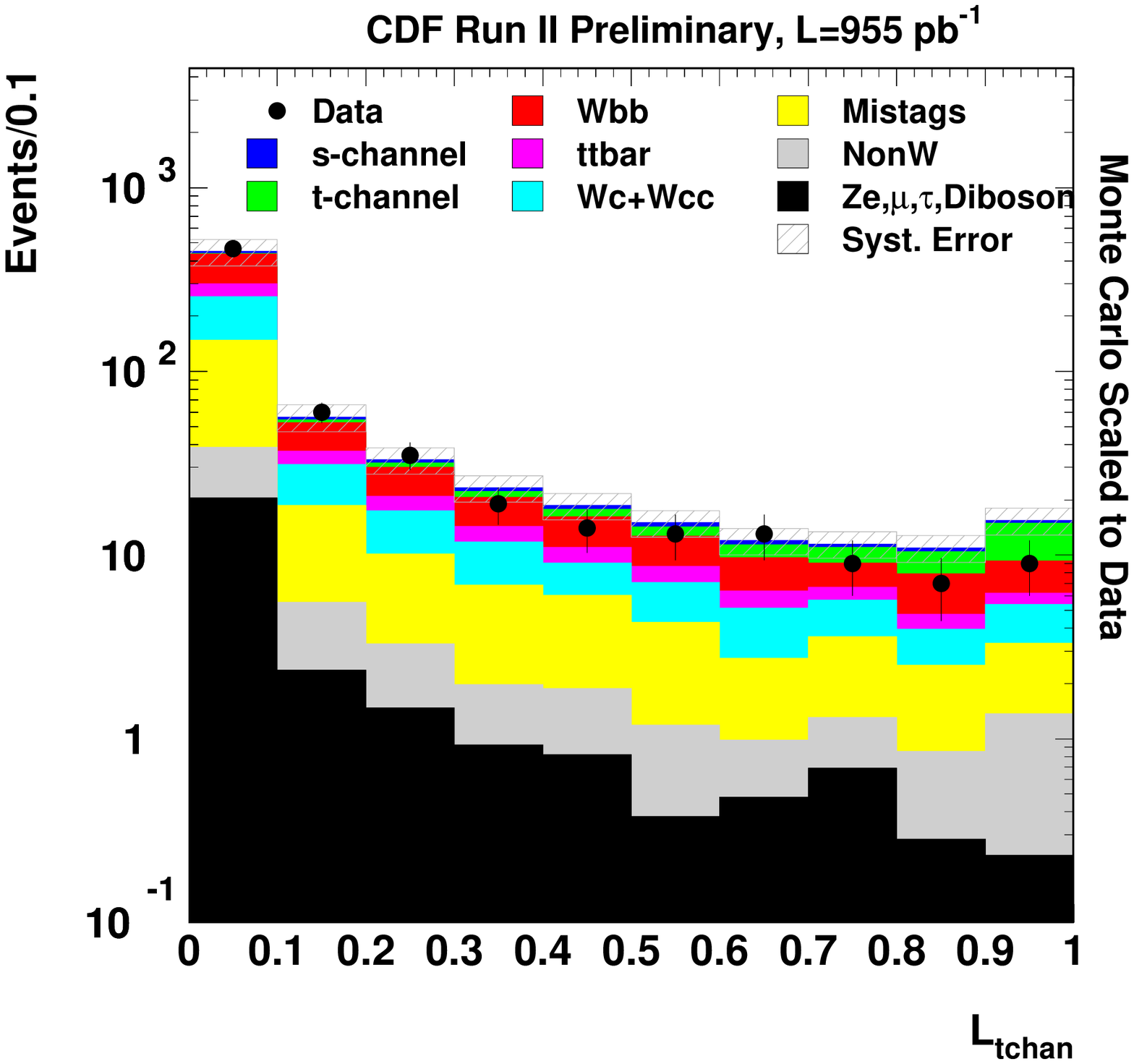}
\end{center}
\caption[nnbtagAndLLD]{\label{fig:nnbtagAndLLD}(a) Output distribution of the 
neural net $b$ tagger for 644 candidate events in the $W+2$ jets bin.
Overlayed are the fitted components of beauty-like, charm-like
and mistag templates The yellow error band indicates the statistical
uncertainties of the fitted sum.
(b) The distributions of the $t$-channel likelihood function for CDF data 
compared to the Monte Carlo distribution normalized to the expected 
contributions. A logarithmic scale is used.} 
\end{figure}

One of the variables used in the analysis is the output of a neural 
net $b$ tagger.
In figure~\ref{fig:nnbtagAndLLD}a the distribution of this $b$ tag
variable is shown for the 644 data events. In case of double-tagged
events the leading $b$ jet (highest in $E_T$) is included in this
distribution.
The neural net $b$ tagger gives an additional handle to reduce the
large background components where no real $b$ quarks are contained,
mistags and charm-backgrounds. Both of them amount to about 50\%
in the $W+2$ jets data sample even after imposing the requirement that one jet
is identified by the secondary vertex tagger of CDF~\cite{secvtx}.

The $t$-channel likelihood function is shown in figure~\ref{fig:nnbtagAndLLD}b.
The best sensitivity (expected p-value 2.5\%) is reached by combining 
the two likelihood discriminants
in a two-dimensional fit where $t$-channel and $s$-channel are considered
as one single-top signal (combined search). 
The observed data show no indication of a single-top signal and are 
compatible with a background-only hypothesis (p-value 58.5\%). 
The upper limit on the combined single-top cross section is
found to be 2.7 pb at the 95\% C.L., while the expected limit is
2.9 pb. The best fit for the cross sections yields
$\sigma_{t}=0.2^{+0.9}_{-0.2}\,\mathrm{pb}$
and $\sigma_{s}=0.1^{+0.7}_{-0.1}\,\mathrm{pb}$.

\subsection{Neural Network Search}
\label{sec:nn}
\begin{figure}[!th]  
\hspace*{0.03\textwidth} {\small (a)}\\
\begin{center}
\includegraphics[width=0.40\textwidth]{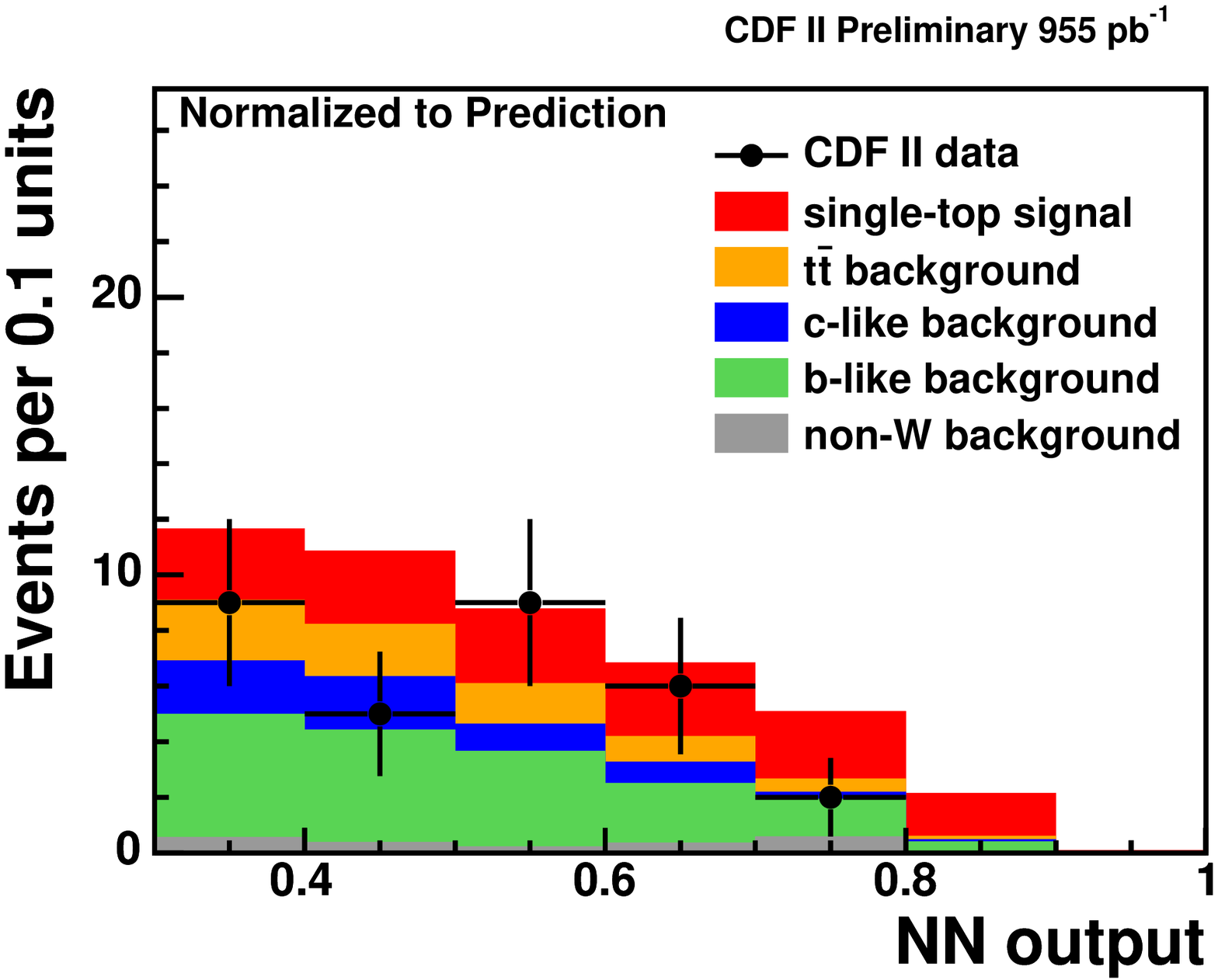}\\
\end{center}
\hspace*{0.03\textwidth} {\small (b)}\\
\begin{center}
\includegraphics[width=0.40\textwidth]{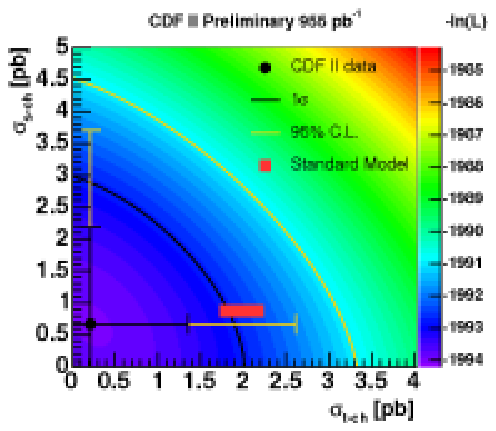}
\end{center}
\caption[nnstdata]{\label{fig:nnstdata}Single-top search with 
  neural networks at CDF:
  (a) Data compared to the standard model expectation in the 
  signal region (neural network outputs larger than 0.3).
  (b) Likelihood contours of the separate neural network search.}
\end{figure}

In the second analysis a neural network is used to combine 23 kinematic or 
event shape variables to a powerful discriminant. The output is given between -1 for background-like events and +1 for signal-like events. The five most important input variables are reported by the neural network package.
They are: the output of the jet-flavor separator neural network, $M_{l\nu b}$, the dijet mass, $\eta$ of the untagged jet times the charge of the isolated lepton and the multiplicity of soft jets in the event with 8 GeV < $E_T$ < 15 GeV. Figure~\ref{fig:nnstdata}a shows the observed data for the combined search
compared to the expectation in the signal region defined by a neural
network output between 0.3 and 1.
In the combined search where the ratio of $t$-channel and $s$-channel 
cross sections is fixed to the standard model value a p-value of
54.6\% is observed, providing no evidence for single-top production.
The corresponding upper limit on the cross section is
$2.6\,\mathrm{pb}$ at the 95\% C.L..
To separate $t$- and $s$-channel production two additional networks are
trained and a simultaneous fit to both discriminants is performed.
The best fit values are $\sigma_t = 0.2^{+1.1}_{-0.2}\,\mathrm{pb}$
for the $t$-channel and $\sigma_s = 0.7^{+1.5}_{-0.7}\,\mathrm{pb}$.
The corresponding upper limits are 2.6 pb and 3.7 pb, respectively.
The observed p-value is 21.9\%.

\subsection{Matrix Element Analysis} 
\label{sec:ME}
\begin{figure}[tbh]

\hspace*{0.03\textwidth} {\small (a)}\\
\begin{center}
\includegraphics[width=0.40\textwidth]{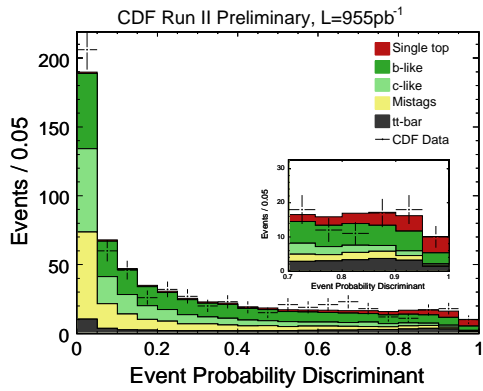}\\
\end{center}
\hspace*{0.03\textwidth} {\small (b)}\\
\begin{center}
\includegraphics[width=0.40\textwidth]{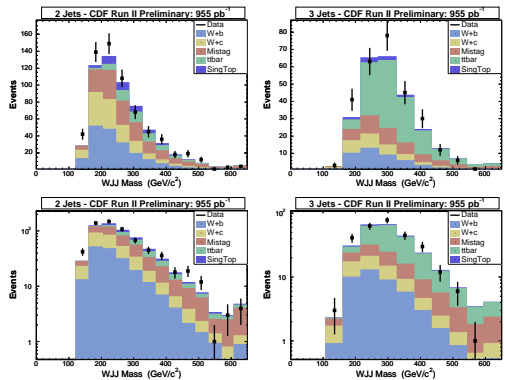}
\end{center}
\caption{\label{fig:meAndWprime}  (a) Event probability density (EPD) based 
  on matrix elements. The 
  inset shows the signal region with EPD $>$ 0.7. 
  (b) $W^\prime \rightarrow t\bar{b}$ search: The invariant mass 
  of the charged lepton, the reconstructed neutrino and the 
  two leading jets.}
\end{figure}

Another way to discriminate signal from background is to compute 
leading order matrix elements for signal and background processes.
The measured four-vectors of the jets and the charged lepton are
used as experimental input to compute event-by-event probability densities.
Constraints on energy and momentum 
conservation are applied. The jet energy measurements are corrected
to parton level energies using transfer functions. To obtain a 
relative weight for a certain hypothesis one integrates over jet
energies and the momenta of the incoming partons using the parton
density functions as integration kernel. The weights for the individual
hypotheses are combined event-by-event by forming a ratio signal
and signal-plus-background weights. The resulting discriminant is named
event probability density (EPD). The measured EPD distribution is 
shown in figure~\ref{fig:meAndWprime}, compared to the fitted event
rates for the different processes.
A single-top signal corresponding to a $2.3\,\sigma$ excess is observed.
The associated single-top cross section is 
$2.7^{+1.5}_{-1.3}\,\mathrm{pb}$.

\section{Search for a $W^\prime$ boson}
Based on a the same event selection as the standard model searches
CDF has also searched for a $W^\prime$ boson in the
decay channel, $W^\prime\rightarrow t\bar{b}$ that appear in models with left-right symmetry, extra dimensions, Little Higgs, and topcolor~\cite{wprime}. 
The signal is modeled using the event generator 
{\sc pythia}~\cite{pythia}.
The invariant mass of the charged lepton, the reconstructed neutrino and the 
two leading jets is used as a discriminant.
The measured data are shown in comparison to the standard model
prediction in figure~\ref{fig:meAndWprime}b. 
No evidence for a resonant $W^\prime$ boson production is found,
yielding limits on the production cross section ranging from
2.3 pb at $M(W^\prime)=300\,\mathrm{GeV}$ to 0.4 pb at  
$M(W^\prime)=950\,\mathrm{GeV}$. Utilizing theoretical cross section
calculations~\cite{wprime},
lower limits on the $W^\prime$ mass are set: 
$M(W^\prime) > 760\,\mathrm{GeV}$ if the mass of potential right-handed
neutrinos is below $M(W^\prime)$ and
$M(W^\prime) > 790\,\mathrm{GeV}$ otherwise.

\section{Conclusions}
The collaboration has performed searches for standard and non-standard model
single-top production using a data set corresponding to an integrated 
luminosity of $955\,\mathrm{pb^{-1}}$.
No evidence for these processes could be established. 
Two standard model searches based on likelihood discriminants or neural
networks find no excess which can be attributed to single-top production,
while the matrix element analysis finds an excess of $2.3\,\sigma$ compatible
with single-top production. The overall consistency of all there analysis is
only 1\%, since the analyses feature a correlation between 60\% and 70\%.
At present, there are no hints to other causes than statistical fluctuations.
Recent measurements utilizing the matrix element method and likelihood discriminant 
using 1.5 $fb^{-1}$ show a 3 $\sigma$ evidence for single top production.


The $W^\prime$ search in the single-top channel
establishes new upper limits of $M(W^\prime) > 760\,\mathrm{GeV}$
or $M(W^\prime) > 790\,\mathrm{GeV}$ depending on the mass of the right-handed 
neutrino.

%
%

\end{document}